**A sustained oscillation in a toy-model of the coupled atmosphere-ocean system**

Oliver Bothe[1]

[1] Klimacampus, CliSAP, Universität Hamburg, c/o Max Planck Institut für Meteorologie, Bundesstraße 53, 20146 Hamburg, Germany

* Correspondence to:

Oliver Bothe

Klimacampus, CliSAP, Universität Hamburg, c/o Max Planck Institut für Meteorologie, Bundesstraße 53, 20146 Hamburg, Germany

e-mail: oliver.bothe@zmaw.de

Phone: ++49 40 41173 371

**Abstract**

Interaction between atmospheric mid-latitude flow and wind-driven ocean circulation is studied coupling two idealized low-order spectral models. The barotropic Charney-DeVore model with three components simulates a bimodal mid-latitude atmospheric circulation in a channel with two stable flow patterns induced by topography. The wind-driven ocean double gyre circulation in a square basin (of half the channel length) is modeled by an equivalent barotropic formulation of the Veronis model with 21 components, which captures Rossby-wave dynamics and nonlinear decadal variability. When coupled, the atmosphere forces the ocean by wind-stress while, simultaneously, the ocean affects the atmosphere by thermal forcing in terms of a vorticity source. Coupled atmosphere-ocean simulations show two stable flow patterns associated with the topographically induced atmospheric bimodality and a sustained oscillation due to interaction between atmospheric bimodality and oceanic Rossby dynamics. The oscillation is of inter-annual to inter-decadal periodicity and occurs in a reasonably wide parameter domain.

# 1. Introduction

The interaction of ocean and atmosphere in the mid-latitudes has been studied in depth throughout the last century. General circulation models (GCM) and simplified models of coupled and driven systems are used to identify mechanisms of mid-latitudinal inter-annual and inter-decadal variability of the climate system. Intrinsic atmospheric variability is responsible for a wide range of observed variations, but there is evidence for (e.g. Ferrari and Cessi 2003) and against (e.g. Kravtsov and Robertson 2002) an active involvement of the ocean in the mid-latitude climate system.

The respective roles of atmosphere and ocean are important in the discussion of possible multi-annual variability of the North Atlantic Oscillation (NAO) (Wallace 2000). The NAO influences the whole Northern Hemisphere among others due to shifts of the storm tracks (Marshall et al. 2001). A relationship is found between the seasonal NAO signature and intra-seasonal variability of blocking episodes (Marshall et al. 2001). Considering the dynamics of blocking in topographically and thermally disturbed flow, Charney and DeVore (1979) display the possible existence of multiple steady-states in mid-latitudinal atmospheric flow in a low-order barotropic channel model – a blocking mode and a zonal mode.

For low frequency variations of the thermohaline and wind-driven ocean circulation numerous studies describe modes on inter-annual to inter-decadal time scales and the importance of internal nonlinear interaction (Jiang et al. 1995; Speich et al. 1995; Simonnet and Dijkstra 2002; Hogg et al. 2005). Less attention has been given to externally induced variability of the wind-driven ocean circulation in a quasi-linear system with minimized nonlinearities. There the only relevant mechanism is the excitation of Rossby waves. Goodman and Marshall (1999) report coupled Rossby waves in a simple coupled atmosphere-ocean model consisting of linear quasi-geostrophic formulations for ocean and atmosphere. Kravtsov et al. (2006, 2007) find in their highly nonlinear coupled model, besides a mechanism of decadal mid-latitude climate variability, an inter-annual mode, which they identify as a coupled Rossby wave as described by Goodman and Marshall (1999). Veronis (1963) introduced a homogenous formulation of the wind-driven single-gyre ocean circulation with a limited number of Fourier modes. His model captures the broad characteristics of the wind-driven problem. Limitations are the reduced number of modes and the expansion of the stream-function into a double-sine series.

A coupled mode of North Atlantic interaction between atmospheric and oceanic circulation is described by Bjerknes (1964). Besides an out-of-phase relationship of atmospheric westerlies and sea temperature anomalies his work hints to a mechanism of coupled low frequency variability due to different adjustment time scales of atmospheric (fast adjustment) and oceanic (decadal adjustment) circulations.

The goal of the present work is to describe a potential mechanism of low frequency climate variability in a low order (that is conceptual) model framework. The focus is on the interaction between atmospheric circulation regimes and the wind-driven ocean circulation. For this purpose a coupled system is developed based on the spectral models for the atmosphere (Charney and DeVore 1979) and the ocean (Veronis 1963). Section 2 specifies the models and their coupling mechanisms. Section 3 describes results and analyses sensitivity to relevant parameters. Results are summarized and discussed in Section 4.

## 2. Atmosphere and ocean models: Description and coupling

Two low-order systems of idealized atmospheric and ocean dynamics in spectral form have been introduced more than a quarter of a century ago based on the quasi-geostrophic, barotropic vorticity equation on a beta-plane: (1) for the atmosphere, a mid-latitude beta-plane channel with topography is formulated by Charney and DeVore (1979), (2) for the ocean, a homogenous barotropic spectral model is derived for a square basin on a beta-plane with atmospheric forcing induced by the curl of wind-stress (e.g. Veronis 1963), which is upgraded here using an equivalent barotropic version. Both models are described in the Appendix.

Coupling the two models requires vorticity sources which are induced, for the ocean, by the curl of atmospheric wind-stress and, for the atmosphere, by thermal forcing associated with the ocean circulation. In addition, the atmospheric flow is forced by a zonal wavenumber-1 mountain-valley configuration with the square ocean being positioned in the valley, where the deepest point is centered at $x = \pi$, $y = \pi/2$, while the mountain-peak is $\pi$ out of phase. A brief introduction to models and coupling follows. Both models compute the stream function. In the introduction of each model we omit distinguishing subscripts for reasons of readability.

*a. Spectral atmospheric model*

The low-order Charney-DeVore model (Charney and DeVore 1979; CDV in the following) is based on barotropic quasi-geostrophic flow of depth $H$ over topography $h(x,y)$ in a mid-latitude beta-channel centered at 45° latitude. Channel dimensions are *0*

$\leq y \leq b\pi$, $0 \leq x \leq 2\pi$, with the scale ratio of meridional width to zonal length $b=2L_y/L_x$. The non-dimensional vorticity equation is

$$\frac{\partial}{\partial t}\nabla^2\psi + J(\psi, \nabla^2\psi + h) + \bar{\beta}\frac{\partial \psi}{\partial x} = -C\nabla^2(\psi - \psi^*) \qquad (1)$$

with the stream-function $\psi(x,y,t)$ and the non-dimensional $\beta$-parameter, $\beta$. The Jacobi operator $J$ is $J(a,b) = \partial a/\partial x\, \partial b/\partial y - \partial b/\partial x\, \partial a/\partial y$. The right hand side is the (barotropic equivalent of a thermal) forcing relaxing the flow to a prescribed vorticity or stream-function profile $\psi^*$. In the coupled scenario this is achieved as described in the coupling section (*c*); the damping coefficient $C = O(10^{-1})$ (or timescale $1/C$) incorporates the Ekman layer depth. The stream-function is decomposed into three spectral modes $\psi_1 = Z\varphi_1$, $\psi_2 = V\varphi_2$ and $\psi_3 = W\varphi_3$ using the spectral expansions $\varphi_1 = 2^{1/2}\cos(yb^{-1})$, $\varphi_2 = 2\cos(x)\sin(yb^{-1})$ and $\varphi_3 = 2\sin(x)\sin(yb^{-1})$. The non-dimensional stream-function forcing $\psi^*$ and bottom topography $h$ read

$$h(x,y) = \frac{1}{2}\varphi_2 = \cos(x)\sin\left(\frac{y}{b}\right)$$

$$\psi^* = \psi_0^*\varphi_1 = \psi_0^*\sqrt{2}\cos\left(\frac{y}{b}\right) \qquad (2)$$

Further details are described in Appendix A.

*b. Spectral ocean model*

The spectral wind-driven ocean circulation model (e.g. Veronis 1963; Böning 1986) is based on the equivalent barotropic quasi-geostrophic vorticity equation

(Pedlosky 1987, 1998) for a square basin ($L=L_x=L_y=\pi$) on a beta-plane, including lateral friction:

$$\frac{\partial}{\partial t}(\nabla^2\psi - Bu\,\psi) + \lambda_I J(\psi, \nabla^2\psi - Bu\,\psi) + \hat{\beta}\frac{\partial\psi}{\partial x} = \lambda_M \nabla^4\psi + curl\,\tau \qquad (3)$$

where $\psi$ is the stream-function; $\lambda_I$ characterizes the influence of the nonlinearities; the Burger number is $Bu = (Rd/L)^2$ with the oceanic Rossby radius of deformation $Rd$ (11 $km \leq Rd \leq 150\ km$); $\hat{\beta} = 1$ is a non-dimensional $\beta$-parameter. The ocean is forced by the curl of the wind-stress, $curl\ \tau = \nabla \times \tau$, proportional to the atmospheric wind, and damped by lateral friction with the coefficient $\lambda_M$. Following Veronis (1963), the stream-function $\psi(x,y,t)$ is decomposed into a double-sine Fourier series. In doing so, it is accepted that this stream-function expansion may not be the one best adapted to the wind-driven ocean-gyre circulation (Simonnet and Dijkstra 2002).

The truncation leads to a set of $M \times N$ ordinary differential equations with the highest zonal and meridional wave-numbers $M$, $N$; $M=7$, $N=3$ appear as the lowest truncation representing sufficient details of the wind-driven ocean-gyre-circulation when compared to stream-function contours for Munk's model as presented by Pedlosky (1998). The model is further described in the Appendix B.

*c. Coupling*

The two spectral low order models are coupled by their respective driving terms: the curl of the atmospheric wind-stress forces the ocean, and the atmosphere is driven by the ocean through a stream-function pattern $\psi^*$ or vorticity pattern.

1) Ocean Forcing (wind-stress curl)

The curl of the wind-stress, $curl\ \tau = C_D\ \zeta$, forcing the ocean model, depends on the non-dimensional drag-coefficient $C_D$ and the atmospheric relative vorticity $\zeta = \nabla^2 \psi_{At}$. As the square ocean basin is centered (as orographic valley) within the atmospheric channel, the atmospheric relative vorticity forcing is confined to this zonal section and to the zonally averaged part. Thus the ocean forcing becomes, for even $l$,

$$curl\ \tau_{k,l} = -C_D \frac{4}{\pi} \sum_{k=1}^{K-1} \frac{1}{k}\left(V\frac{4}{\pi}\left(1+\frac{1}{b^2}\right) - Z\frac{2\sqrt{2}}{b^2} \sum_{l=2}^{L} \frac{1}{l^2-1}\right)$$

(4)

2) Atmospheric forcing (vorticity source)

The CDV atmosphere is forced by Newtonian relaxation towards a prescribed meridional vorticity profile $\psi^*$, which is zonally symmetric (Eq. 1, right hand side). Interpreting the meridional vorticity gradient as the barotropic equivalent of thermal forcing, it can be induced in terms of a (sea surface) temperature gradient associated with the modeled ocean circulation. The anomaly of the sea surface temperature provides a time varying lower boundary condition for the atmospheric model. Thus a time-dependent expression for the ocean (or sea surface) temperature is required. To obtain an expression for the atmospheric vorticity source ($\psi^*$), a low order system is introduced for the temperature equation, consisting of only two components representing the zonal mean $T_0$ and its anomaly $T_1$. The form of the temperature field is motivated by Cessi (2000) and van der Avoird et al. (2002). Starting from

$$\frac{\partial T}{\partial t} = -J(\psi_{Oc}, T) + r_c(\theta - T) + \varepsilon \nabla_h^2 T \tag{5}$$

and projecting temperature $T$ and its equilibrium $\theta$ onto basis functions, $T = T_0 \cos(y) + T_1 \cos(x) \sin(y)$ and $\theta = \theta_e \cos(y)$, yields a system of two coupled equations for the temperature:

$$\dot{T}_0 = r_c a_r (\theta_e - T_0) - \varepsilon T_0 - \frac{2}{3}\psi_{1,1} T_1$$

$$\dot{T}_1 = -r_c T_1 - 2\varepsilon T_1 + \frac{4}{3}\psi_{1,1} T_0 \tag{6}$$

with the coefficient for relaxation $r_c$ and diffusion term $\varepsilon = \lambda_M$, and the ocean stream-function $\psi_{Oc} = \psi_{1,1}$. For the ocean-to-atmosphere coupling a wavenumber-1 ($r=s=1$, see Appendix B) ocean flow field suffices which is interpreted as the anomaly of a western boundary current. Following Gallego and Cessi (2000), the relaxation of the zonal mean temperature requires a correction $a_r < 1$, when applying the temperature as the atmospheric forcing, so that $T_0$ and $T_1$ relax on different time scales. Using only the zonally averaged meridional gradient $T_0$, the atmospheric stream-function forcing $\psi^*$ becomes $\psi^* = \psi^* T_0 = 1/2^{1/2} C_T T_0$; $C_T$ is the non-dimensional coefficient for scaling.

In summarizing, the atmosphere-ocean coupling can be described, in qualitative terms, by sea surface temperature anomalies changing the meridional location and the intensity of the atmospheric jet which, in turn, affects the ocean by modifying the curl of the wind-stress. This leads to a change of the temperature gradient and a variation of the atmospheric vorticity source. Simulations with the coupled Charney-DeVore and modified Veronis models are performed using the classical explicit 4-th order Runge-

Kutta scheme. The coupled runs start from rest, so the initial forcing or tendency is given solely by the temperature relaxation term. The parameter constellation used can be found in table 1.

## 3. Results

The dynamics of the uncoupled Charney-DeVore model are represented by two stable equilibria without internal variability for a wide range of the parameters $C$, $\psi^*$ and $\beta$ which characterize the Ekman damping, the forcing strength and the $\beta$-term respectively (Charney and DeVore 1979). Figure 1 (left) displays the stream-functions of the equilibrium solutions: Equilibrium-1 describes a zonal flow state, equilibrium-2 shows a strong wave component or blocking state. The latter is characterized by a low about $\pi/4$ east of the top of the mountain and a blocking high about $\pi/4$ east of the centre of the channel. The modified Veronis model responds with double (sub-polar/sub-tropical) gyre circulations (Figure 1, right) when prescribing the two atmospheric equilibria to obtain the required wind-stress forcing. The terms 'sub-polar' and 'sub-tropical' do not relate to the real ocean circulation but are used as reference to the simulated cyclonic (sub-polar) and anti-cyclonic (sub-tropical) gyres. While the zonal flow state yields a stronger ocean circulation with small asymmetry, the blocking state generates a rather diminished sub-polar gyre. The low-order ocean model reveals a reasonable approximation of the wind-driven gyre circulation (e.g. Pedlosky 1998). Nonlinear internal interactions in the ocean model are reduced due to the setup ($\lambda_I$ small in Eq. 2).

*a. Coupled models*

Coupling the atmosphere-ocean-system displays different types of solutions. Dependent on the parameter constellations steady-states, which are characterized by the two atmospheric equilibrium states constantly forcing the ocean, and a sustained oscillation are found. Phenomenology and underlying mechanism of the latter are discussed in the following (for the parameter constellation displayed in table 1).

1) Phenomenology

The oscillation is characterized by transitions between a weak and a strong zonal flow in the atmosphere with an inter-annual time scale of about four years, during which the atmospheric stream-function-field visits the neighborhood of the two uncoupled equilibrium states without a state of rest. Figure 2a-d displays the evolution of the atmospheric and oceanic zonal mean stream-functions and their anomalies through one oscillation period. Based on the atmospheric flow patterns one cycle can be sub-divided into four segments (marked by I, II, III and IV in Figure 2): Two slow transitions (segments I and III) and two rapid changes (segments II and IV). While the transition phases last up to several years dependent on the parameters used, the rapid changes occur within a few months. During transition I, a continuous reduction of the atmospheric stream-function gradient (or the strength of the atmospheric zonal flow) is connected with an enhanced oceanic double gyre due to the latter's delayed reaction to the atmospheric forcing. While the sub-tropical gyre obtains its maximum strength in the middle of segment I, the sub-polar gyre increases until the rapid change II. The strengthening of the oceanic circulation is connected with a decrease of the oceanic temperature gradient. At rapid change II, this gradient (which forces the atmosphere) attains its minimum (Figure 2e) and the atmospheric circulation switches into the state with a strong wave component and a weak stream-function gradient. Then, in transition phase III, the atmosphere relaxes towards a stronger zonal flow. The ocean is characterized by a reduced gyre circulation. The minimum oceanic circulation is obtained at the rapid change IV. Both gyres reach their minimum almost simultaneously, when the temperature gradient attains its maximum. The atmosphere shows a rapid switch into a zonal flow state, and phase I of the oscillation starts again.

The average of the atmospheric circulation over one full oscillation cycle shows a low about a quarter wavelength downstream of the mountain and a high upstream (Figure 3a), while the time mean ocean circulation exhibits stronger flow in the sub-polar than in the sub-tropical gyre (Figure 3b).

2) Mechanism

The phase space trajectory of the CDV-model during one cycle of the oscillation is displayed in figure 4 with the system oscillating about the neighborhoods of the two atmospheric fixed points. Using a nonlinear CDV-model of six components, Crommelin et al. (2004) associate such behavior with preferred paths of regime transition in the atmosphere and relate them to heteroclinic cycles. Based on the phase space trajectory the first slow transition (phase I) drives the atmospheric model away from the vicinity of the zonal flow equilibrium towards the attractor region of the 'blocked' regime. Meanwhile the oceanic gyre circulation increases (the temperature gradient decreases) as delayed response to the strong atmospheric flow (see figure 2e). The temperature tendency reversal, which is due to the changed atmospheric conditions, emerges just before the atmospheric model approaches the immediate neighborhood of the weak zonal flow equilibrium during the rapid change II; that is some days after a discernable change of atmospheric flow is observed. The resulting temperature increase is too strong to allow the atmospheric model to reach the weak zonal flow steady state. Thus the atmosphere oscillates about it and then is forced into a transition back towards the strong zonal flow regime (phase III). During this transition the temperature gradient reaches its maximum and decreases in the following, such that the atmospheric model again cannot adopt a steady state (phase IV). Instead, it oscillates in the phase space sector of the equilibrium and then proceeds into a slow relaxation (phase I).

Slow transition and rapid change segments are clearly separated by their respective time scale. The ocean determines the time scale of slow evolution by its delayed response to the atmospheric forcing while the spontaneous regime transitions of the CDV-model shape the coupled atmosphere-ocean-system's behavior during the rapid changes. The nearly quadrennial timescale of the observed oscillation matches principally the long Rossby wave transit time for the studied ocean basin size. This is in agreement with Cessi and Primeau (2001); there the variation of the wind-stress excites oceanic Rossby waves which, in turn, influence the atmospheric vorticity source and thereby intensity and position of the maximum of atmospheric flow.

In summarizing, the observed low frequency variability evolves due to the following mechanism: (i) Changing wind-stress curl excites Rossby wave like anomalies in the ocean. These anomalies lead to a slow change of the zonally averaged temperature forcing of the atmosphere, and, finally, (ii) to a rapid shift of the preferred atmospheric flow regime, (iii) which, in turn, causes a slow transition due to oceanic Rossby wave dynamics; (iv) the cycle is completed by a second rapid transition to the other atmospheric circulation regime.

*b. Sensitivity*

Two characteristics of the coupled atmosphere-ocean-model appear to control the obtained results: (i) the topographically induced bimodality of the CDV-model allowing fast regime transitions and (ii) the dynamics of oceanic Rossby waves setting the time scale of slow temperature trends. In the following the sensitivity to these characteristics is investigated by varying (i) the atmospheric relaxation coefficient $C$ (incorporating the atmospheric Ekman layer depth) and (ii) the Burger number $Bu=(Rd/L)^2$, which implicitly corresponds to the Rossby radius of deformation $Rd$.

Figure 5 shows two ($C$, $Rd$)-diagrams of the coupled model oscillation regimes with respect to varying damping coefficient, $C$, and varying Rossby radius of deformation, $Rd$. Panel (a) of Figure 5 displays the behavior of the oscillation period $T_{osci}$ dependent on $C$ and $Rd$. Shading shows the transition time in years. For blank areas, steady states are obtained or the oscillation period is shorter than about one month. In panel (b) the oscillation time $T_{osci}$ has been normalized by the theoretical transit time of the longest Rossby wave, $T_{Rd} = L_x\ Rd^{-2}\ \beta^{-1}$; blank areas denote steady states or normalized oscillations shorter than ten percent of $T_{Rd}$. In a range of $C$ depending on the Rossby radius the sustained oscillation occurs. Outside this range the atmosphere-ocean system attains a steady state.

1) Relaxation coefficient $C$

The atmospheric Ekman pumping controls the input of momentum to the ocean-model, and combined with the forcing strength $\psi^*$, the existence of one or several steady-states of the CDV-model. Higher values of $C$ mean the ocean receives stronger forcing and, contrasting this, the relaxation time scale, $1/C$, shortens. The sustained oscillation occurs for a wide range of Ekman pumping. Within this *C*-interval, as the coefficient is enlarged, the oscillation appears in a smaller subspace of the atmospheric phase space and its characteristics are altered. For low and for large *C*-values the atmosphere-ocean-model approaches steady-state solutions. For large values of the relaxation coefficient *C* the steady state is approximately located in the middle between the 'blocked' and zonal equilibrium states of the CDV-model, while for weak Ekman pumping (low *C*-values) the model adopts one of these two equilibria (zonal flow state for *C* very small, 'blocked' state for slightly larger values).

2) Rossby radius of deformation *Rd*

The Rossby radius of deformation *Rd* influences the time scale of Rossby wave disturbances. Chelton et al. (1998) give a range of *Rd* for the ocean of about 10 km in the high latitudes to about 230 km at the equator. The model incorporates *Rd* via the Burger number *Bu*, which directly influences its dynamics. Different Rossby wave basin crossing times correspond to different advection times of the temperature equation, thus changing the tendency of the atmospheric forcing. Values between 11 km and 150 km are used for *Rd*. Indeed, changing *Rd* leads primarily to different periods of the sustained oscillation, but high values of *Rd* also induce higher amplitudes of the oscillation and thus alter the characteristics of atmospheric phase space trajectories.

Figure 5b shows the dependence of the observed period depending on *C* and *Rd* normalized by the theoretical transit time of the longest Rossby wave. At $C = 0.2$, corresponding to an Ekman layer depth of $D_e = 200$ m and an atmospheric relaxation time $1/C = 11.6$ days, modeled periods differ about 20% from theoretically calculated transition times for $Rd < 40$ km, while an intermediate range of $Rd$ (40 km $\leq Rd <$ 80 km) leads the coupled system to a steady-state with a strong zonal atmospheric flow. For slightly higher values the model oscillates about this state and for even larger values ($Rd > 120$ km) again a good relationship to the theoretical transfer time is achieved. Stationary solutions for medium range radii of deformation do not occur for larger values of *C*. As to be seen in figure 5b, medium ranged Rossby radii still show an irregular behavior. While for most values of $Rd$ the period of the oscillation decreases continuously with *C* increasing, for 60 km $< Rd <$ 90 km periods as well as amplitudes fluctuate.

Coupled periods mostly differ from the theoretical transit time and the deviation increases with enlarging *C* until a steady state is attained. At $Rd = 37.8$ km the periods of oscillatory solutions are reduced to approximately 1 year for $C = 0.26$ or about 25% of the theoretical transit time. This is accompanied by a reduction of the passed phase space volume. Visualizing the two CDV steady-states to be deterministic potentials (Sura 2002), the system for larger Ekman layer depths does not reach as far into the potential well. This is achieved by smaller values of the effective forcing depending on the state of the ocean. Shortening the relaxation time scale, $1/C$, prevents the ocean from changing the atmospheric equilibrium, the tendencies nevertheless suffice to induce an oscillation about the 'blocked' atmospheric state; these shorter periods still depend on the longest theoretical Rossby wave transit time.

Obviously the influence of the atmosphere in the coupled system becomes more important for *(i)* long relaxation time scales and weak input of momentum to the ocean-model with *C* small, and *(ii)* short relaxation time scales and strong input with *C* large. Then the ocean's delayed reaction to the atmospheric forcing prohibits an oscillation for all values of the Rossby radius of deformation, because the atmospheric vorticity source is too weak or too fast changing. On the other hand the ocean's dynamics characterizes the coupled system for medium sized values of *C*.

In summarizing, the coupled atmosphere-ocean-model shows a sustained oscillation for a wide range of parameters. This coupled mode is shaped by characteristics of the sub-models: (i) multiple steady states in the Charney-DeVore model controlled by the thermal (vorticity) forcing and (ii) oceanic Rossby waves in the modified Veronis model controlled by the Rossby radius of deformation.

**4. Summary and discussion**

Mechanisms of low frequency variability in the atmosphere-ocean-system are investigated coupling two classical spectral low-order models of the wind-driven ocean circulation and the atmospheric mid-latitude flow by their respective driving terms, the wind-stress curl and the barotropic equivalent of a thermal vorticity source. The models are an equivalent barotropic double-gyre formulation of the wind-driven ocean circulation (following Veronis 1963) and a three component version of the Charney-DeVore (1979) model, both formulated on a beta-plane. As far as known to the authors, these spectral models are coupled for the first time. This conceptual low-order – or toy – model setup and its dynamics is the centre of this study.

The simple coupling strategy (the curl of wind-stress forcing the ocean, while the oceans thermal response acts on the atmosphere) results in an asymmetric oceanic double gyre with a larger cyclonic (sub-polar) circulation gyre. The thermal coupling of the ocean to the atmosphere does not include all possible influences affecting the north-south temperature gradient and thus the atmosphere. Furthermore, the expansion of the equivalent barotropic vorticity equation into a double-sine-series alters the characteristics of the oceanic dynamics.

For a wide range of parameters the interaction between both models excites a sustained oscillation of the coupled atmosphere-ocean-system. Its behavior is shaped by the low-order characteristics of the CDV-model and its period is proportional to the transit-time of the longest oceanic Rossby mode. The underlying mechanism is as follows: Changing atmospheric wind-stress leads to a basinwide oceanic Rossby wave anomaly. This anomaly influences the forcing of the atmospheric zonal flow hence driving the atmosphere from the neighborhood of one steady state to the other passing the unstable one. The variability of the atmospheric zonal flow (atmospheric jet position and intensity) sustains the basinwide oceanic mode, that is, a double gyre anomaly. This mechanism is similar to ocean-only calculations of Sura et al. (2000) and simplified model results by Gallego and Cessi (2000, 2001).

The toy-model employed in this study describes atmosphere-ocean interactions in absence of weather 'noise'. Results presented link quasi-linear oceanic Rossby wave dynamics and atmospheric flow regimes to, dependent on the parameter setting, inter-annual to inter-decadal climate system variability. This demonstrates an active oceanic influence on mid-latitude climate variability, when interpreted in the context of Bjerknes' (1964) observations of Atlantic air-sea interaction and of nonlinear simplified studies (e.g. Ferrari and Cessi 2003; van Veen et al. 2001; van Veen 2003). The phase-differences between oceanic stream-function, (sea surface) temperature and atmospheric westerly wind anomalies agree with the findings of Bjerknes (1964) and Gallego and Cessi (2000). Combined interpretation of the coupled Rossby wave (Goodman and

Marshall 1999), its occurrence in a highly nonlinear coupled model (Kravtsov et al. 2006a,b), the ocean only mode found by Ferrari and Cessi (2003) and the present results indicates that coupled Rossby waves are a common feature of atmosphere-ocean interaction and mid-latitude climate variability. This strengthens the concept (e.g. Goodman and Marshall 1999) that the 'ocean "imprints" itself back on the atmosphere on longer timescales'. Although the oscillation found may be unique to the chosen model set-up, preliminary results with a baroclinic spectral low-order model of the atmospheric mid-latitudinal circulation suggest a similar mechanism occurring there.

**Appendices:**

**Appendix A:**

**Charney-DeVore-1979 model**

The low-order Charney-DeVore model (Charney and DeVore 1979; CDV) uses the barotropic quasi-geostrophic equations describing atmospheric flow of height $H$ over topography $h(x,y)$ in a mid-latitude beta-plane channel. The non-dimensional vorticity equation is obtained with a rigid-lid approximation, scaling with a characteristic height $H$, a timescale $\sigma^{-1}$, a horizontal length scale $k^{-1}$, and characteristic amplitude of topography $h_0$ and reads

$$\frac{\partial}{\partial t}\nabla^2\psi + J(\psi, \nabla^2\psi + h) + \bar{\beta}\frac{\partial \psi}{\partial x} = -C\nabla^2(\psi - \psi^*) \qquad (A1)$$

with stream-function $\psi(x,y,t)$, channel dimensions $0 \leq y \leq b\pi$, $0 \leq x \leq 2\pi$ and the non-dimensional aspect ratio of meridional width to zonal length, $b = 2L_y/L_x$. The channel is zonally periodic: $\psi(x,y,t) = \psi(x + 2\pi, y, t)$. The stream-function $\psi$ is constant at the meridional boundaries, $\psi = 0$, $\psi = \pi$, where the ageostrophic boundary condition implies furthermore: $\int \partial \psi / \partial x \, dx = 0$. The term $\bar{\beta} = (\beta \, L_x \, H)/(2 \, \pi \, f_0 \, h_0)$ is the non-dimensional $\beta$-parameter and the Jacobi operator $J$ is $J(a,b) = \partial a/\partial x \, \partial b/\partial y - \partial b/\partial x \, \partial a/\partial y$. The assumed barotropic equivalent of a thermal forcing is found on the right hand side of Eq. (A1) as relaxation to a prescribed vorticity profile $\psi^*$, and $C$ is a relaxation coefficient including the atmospheric Ekman layer depth. The dimensional equations may be found at Pedlosky (1987) and for a more detailed derivation of the non-dimensional system see Sura (2000) and citations there. To obtain a simple spectral model Eq. (A1) is projected onto basis functions which meet the boundary conditions. The CDV-model of lowest dimension consists of the non-dimensional stream-function with three modes, $\psi = \psi_1 + \psi_2 + \psi_3$,

$$\psi_1 = Z\varphi_1 = Z\sqrt{2} \cos\left(\frac{y}{b}\right)$$

$$\psi_2 = V\varphi_2 = V 2 \cos(x) \sin\left(\frac{y}{b}\right) \quad \text{(A2)}$$

$$\psi_3 = W\varphi_3 = W 2 \sin(x) \sin\left(\frac{y}{b}\right)$$

and non-dimensional stream-function forcing $\psi^*$ and bottom topography $h$:

$$h(x,y) = \frac{1}{2}\varphi_2 = \cos(x) \sin\left(\frac{y}{b}\right)$$

$$\psi^* = \psi_0^* \varphi_1 = \psi_0^* \sqrt{2} \cos\left(\frac{y}{b}\right) \quad \text{(A3)}$$

Inserting Eqs. (A2)–(A3) into Eq. (A1) yields the CDV-set of ordinary differential equations:

$$\dot{Z} = \frac{4b^2}{3\pi\sqrt{2}} W - C(Z - \psi^*)$$

$$\dot{V} = -\frac{b^2}{1+b^2}\left(\frac{4}{3\pi\sqrt{2}} ZW - \bar{\beta} W\right) - CV \qquad (A4)$$

$$\dot{W} = \frac{b^2}{1+b^2}\left(\frac{4}{3\pi\sqrt{2}} ZV - \bar{\beta} V - \frac{2}{3\pi\sqrt{2}} Z\right) - CW$$

For more details see also Sura (2002) and Crommelin et al. (2004).

**Appendix B:**

**Spectral ocean model**

The spectral wind-driven ocean circulation model originally presented by Veronis (1963) (further investigated by Böning 1986) is used in an equivalent barotropic formulation of the quasi-geostrophic vorticity equation on a beta-plane, including lateral friction (1.5 layer, that is, one active layer, one infinitely deep resting layer, reduced gravity):

$$\frac{\partial}{\partial t}(\nabla^2 \psi - Bu\,\psi) + \lambda_I J(\psi, \nabla^2 \psi - Bu\,\psi) + \hat{\beta}\frac{\partial \psi}{\partial x} = \lambda_M \nabla^4 \psi + curl\,\tau \qquad (B1)$$

where $\psi$ is the stream-function in a square basin of dimension $L = L_x = L_y = \pi$; $\lambda_I$ and $\lambda_M$ are parameters characterizing the influence of the nonlinearities and the lateral friction respectively. The Burger number is $Bu = (Ro/Fr)^2 = (Rd/L)^2$ with the oceanic Rossby radius of deformation *Rd* (*Ro*, Rossby number, *Fr* Froude number). The atmospheric wind-stress, *curl* τ = ∇ × τ, describes the forcing of the ocean circulation. *ψ* = *0* at *x* = *0,π/a* and *y* = *0,π* and ∇² *ψ* = *0* at *x* = *0,π/a* and *y* = *0,π*, no flux and free-slip conditions apply. Following Veronis (1963), the stream-function *ψ(x,y,t)* is decomposed into a double-sine Fourier series, $\psi(x,y,t) = \sum_{k=1}^{\infty}\sum_{l=1}^{\infty} \Psi_{k,l}(t)\sin(akx)\sin(ly)$ with the spectral coefficient Ψ and the (*k*, *l*)-wave numbers in *x*- and *y*-direction. With the aspect-ratio of a square basin $a = 1$ the spectral coefficients evolve as:

$$\dot{\psi}_{r,s} = -\left(\frac{1}{r^2+s^2-Bu}\right)\left\{-\frac{\lambda_I}{4}\sum_{k=1}^{\infty}\sum_{l=1}^{\infty}\sum_{m=1}^{\infty}\sum_{n=1}^{\infty}\left((m^2-n^2)\psi_{k,l}\psi_{m,n}\Xi\right)\right.$$
$$\left. -\frac{4r}{\pi}\hat{\sum}\left(\frac{m}{r^2-m^2}\psi_{m,s}\right) + \lambda_M(r^2+s^2)^2\psi_{r,2} + curl_{r,s}\tau\right\}$$

(B2)

The hatted $\hat{\sum}$ means summing over odd *m* for even *r* and even *m* for odd *r*. Ξ stems from the Jacobi operator *J* taking the form:

$$\Xi = \text{sgn}(i-k)\text{sgn}(l-j)(-kl+li) \quad \text{for } |i-k| = r \text{ and } |l-j| = s$$
$$+ \text{sgn}(i-k)(-kj-li) \quad \text{for } |i-k| = r \text{ and } |l+j| = s$$
$$+ \text{sgn}(l-j)(-kj-li) \quad \text{for } |i+k| = r \text{ and } |l-j| = s$$
$$+ (-k+li) \quad \text{for } |i+k| = r \text{ and } |l+j| = s \quad \text{(B3)}$$

where *sgn(a - b)* means sign of *a-b*. Truncation leads to a set of $M \times N$ ordinary differential equations with the highest zonal and meridional wave-numbers *M, N*. Wave-numbers *M=7, N=3* appear as the lowest suitable truncation for coupled low-order dynamics which contains sufficient details of ocean dynamics; resolutions have been tested up to *17 × 17*.

**Acknowledgements**

Financial support by SFB512 and CliSAP is acknowledged. Discussion with and advice by Klaus Fraedrich and Frank Lunkeit improved this work. This manuscripts originates from the author's diploma-thesis in meteorology 'Niederfrequente Variabilität in einem niedrigdimensionalen Spektralmodell des Ozeans'.

**Table captions**

Table 1

Values of used non-dimensional parameters for the basic experiment.

**Figure captions**

Fig. 1. Equilibrium solution stream-function patterns of the atmosphere CDV-model (right) and the ocean-model response on the atmospheric forcing (left): a) Equilibrium 1 is a state with predominantly zonal flow yielding a strong double gyre circulation of reduced asymmetry. b) Equilibrium 2 is a blocking state with a prominent wave disturbance forcing a weaker double gyre with a rather weak sub-polar gyre. The orography is shown in light lines. The black bar at the bottom shows the location of the ocean relative to the atmospheric channel.

Fig. 2. Latitude-time diagram (one period) of the (a) oceanic and (b) atmospheric zonal mean stream-functions and their anomalies (c) and (d). (e) Development of the amplitude of the mean temperature coefficient $T_0$ over one period. The four phases of the oscillation are marked as I (first slow relaxation), II (first rapid transition), III (second slow relaxation) and IV (second rapid transition).

Fig. 3. Time mean state of the coupled system for atmosphere (a) and ocean (b).

Fig. 4. Phase space trajectory of the three component Charney-DeVore model over one period for the basic experiment

Fig. 5. (a) Regime diagram of the oscillation period dependent on the Rossby radius of deformation, $Rd$, and the atmospheric relaxation coefficient, $C$, (b) the same for normalized oscillation periods. The period has been normalized by the theoretical transit

time of the longest Rossby wave: $T_{Rd} = L_x \, Rd^{-2} \, \beta^{-1}$. Shading is for different (normalized) transition times. Blank areas in (a) denote steady states or periods shorter than one month or in (b) normalized oscillation periods shorter than ten percent of $T_{Rd}$.

Table 1. Values of used non-dimensional parameters for the basic experiment.

| Atmosphere | Ocean | Coupled system |
|---|---|---|
| $C = 0.2$ | $Bu = 868.79$ | $\theta_e = 2.85$ |
| $\bar{\beta} = 3.61$ | $\hat{\beta} = 1$ | $r_c = 0.64$ |
| $b = 1$ | $\lambda_M = 0.007$ | $\varepsilon = \lambda_M$ |
| | $\lambda_I = 0.0004$ | $a_r = 0.23$ |
| | | $C_D = 0.06$ |
| | | $C_T \approx 6.94$ |

Fig. 1. Equilibrium solution stream-function patterns of the atmosphere CDV-model (right) and the ocean-model response on the atmospheric forcing (left): a) Equilibrium 1 is a state with predominantly zonal flow yielding a strong double gyre circulation of reduced asymmetry. b) Equilibrium 2 is a blocking state with a prominent wave disturbance forcing a weaker double gyre with a rather weak sub-polar gyre. The orography is shown in light lines. The black bar at the bottom shows the location of the ocean relative to the atmospheric channel.

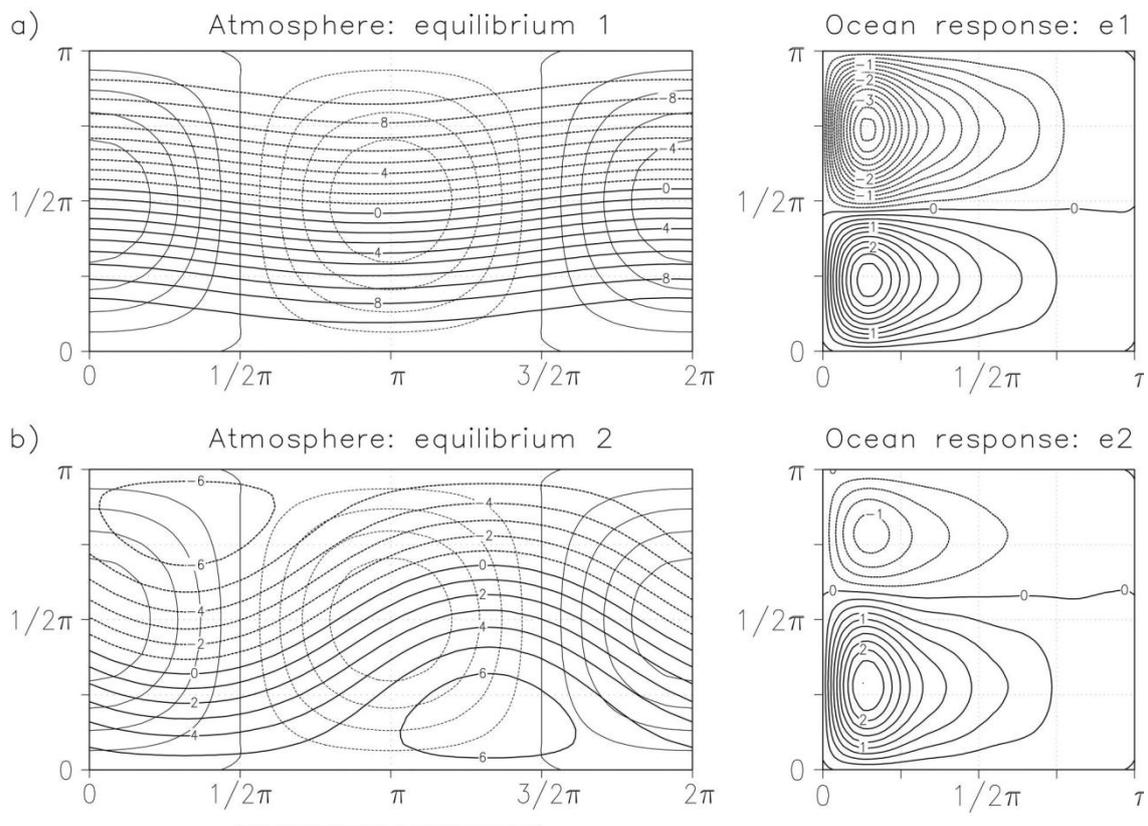

Fig. 2. Latitude-time diagram (one period) of the (a) oceanic and (b) atmospheric zonal mean stream-functions and their anomalies (c) and (d). (e) Development of the amplitude of the mean temperature coefficient $T_0$ over one period. The four phases of the oscillation are marked as I (first slow relaxation), II (first rapid transition), III (second slow relaxation) and IV (second rapid transition).

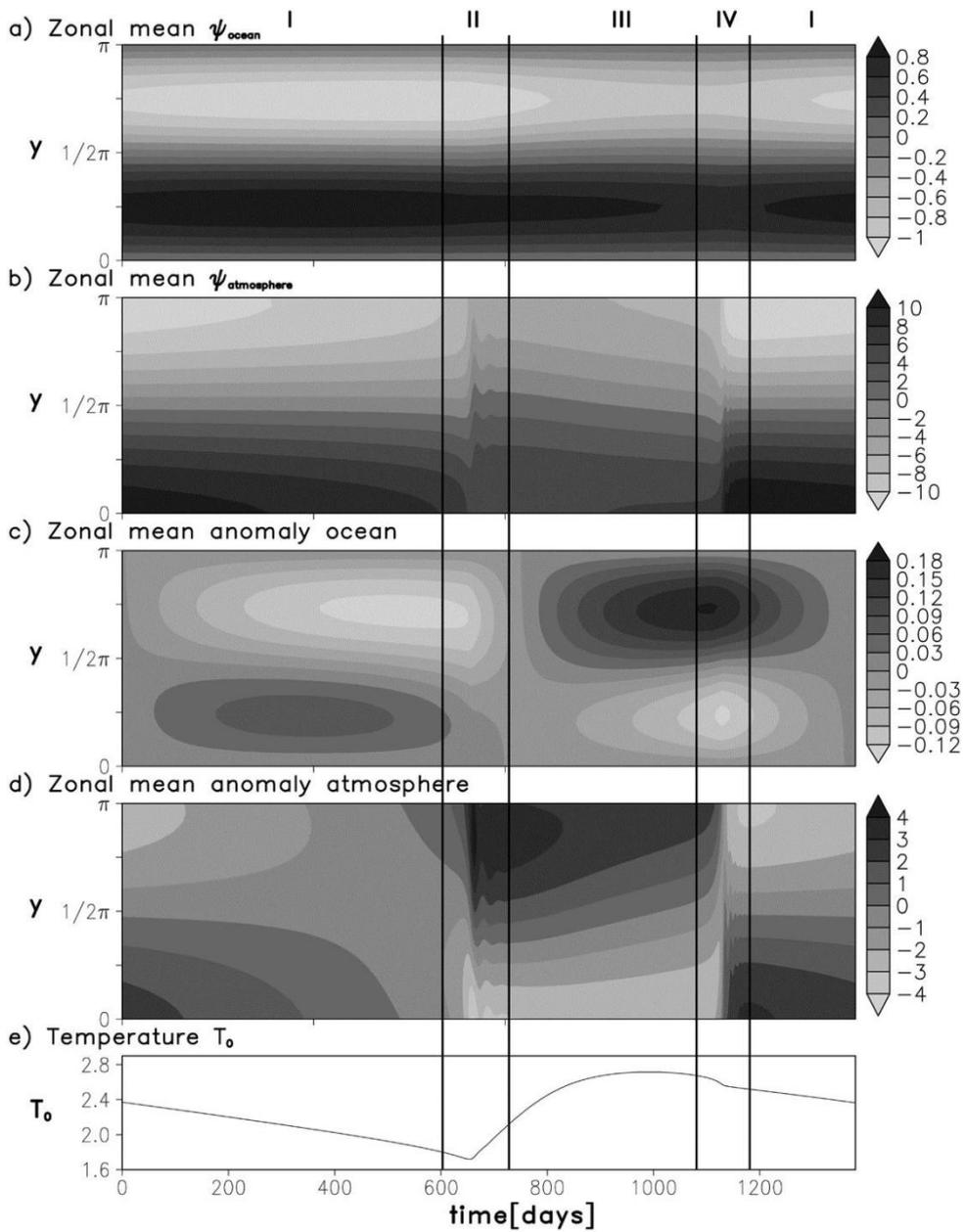

Fig. 3. Time mean state of the coupled system for atmosphere (a) and ocean (b).

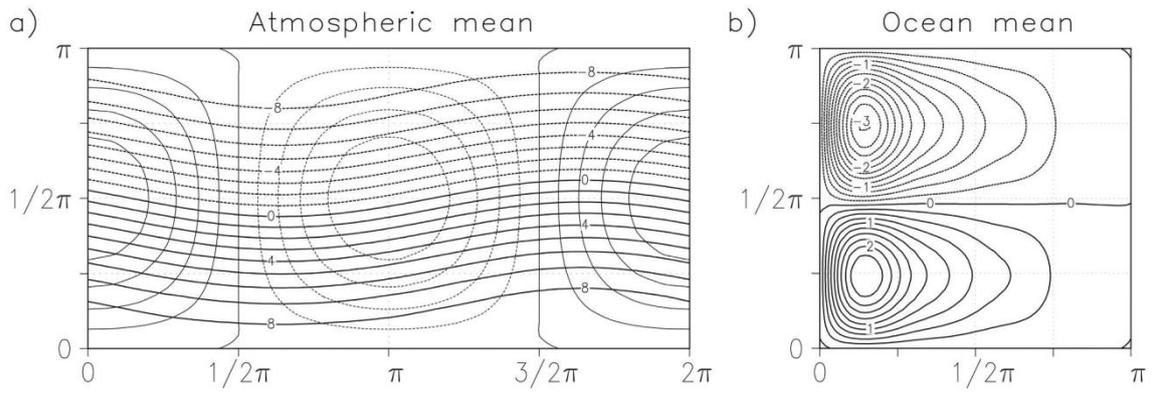

Fig. 4. Phase space trajectory of the three component Charney-DeVore model over one period for the basic experiment

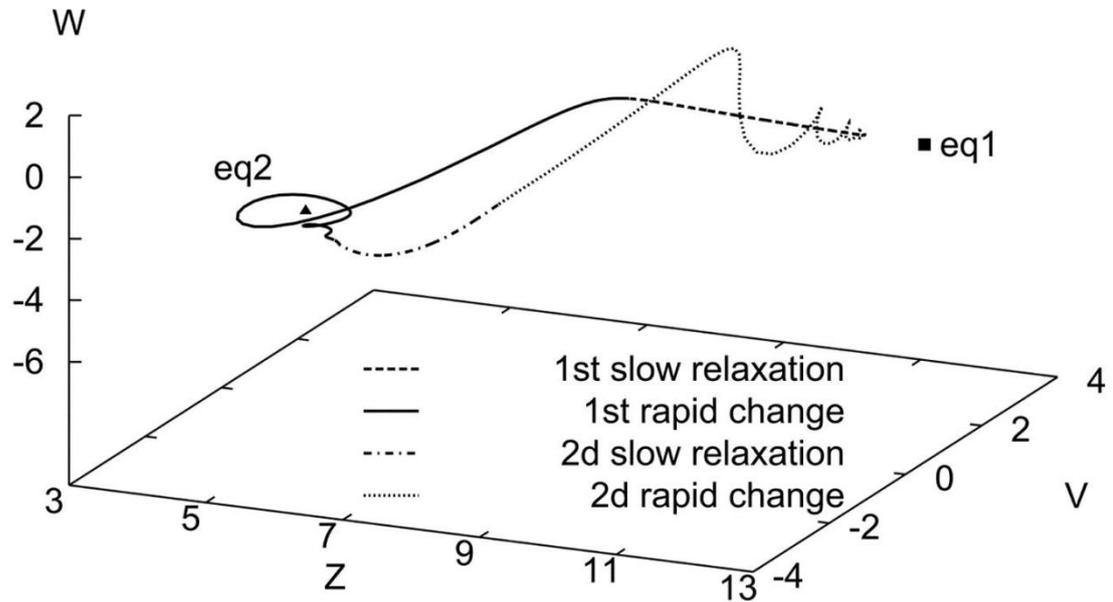

Fig. 5. (a) Regime diagram of the oscillation period dependent on the Rossby radius of deformation, *Rd*, and the atmospheric relaxation coefficient, *C*, (b) the same for normalized oscillation periods. The period has been normalized by the theoretical transit time of the longest Rossby wave: $T_{Rd} = L_x\, Rd^{2}\, \beta^{-1}$. Shading is for different (normalized) transition times. Blank areas in (a) denote steady states or periods shorter than one month or in (b) normalized oscillation periods shorter than ten percent of $T_{Rd}$.

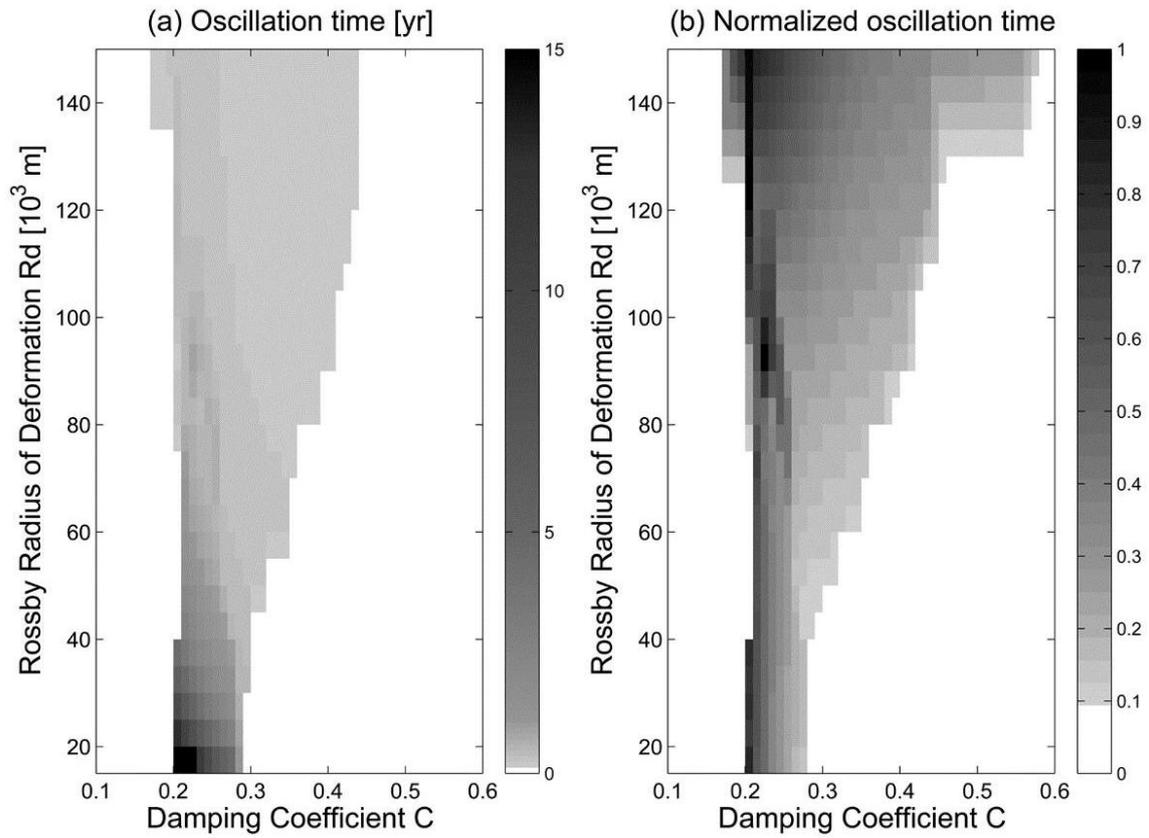